\documentclass[journal]{IEEEtran}
\usepackage{cite}
\usepackage{amsmath,amssymb,amsfonts}
\usepackage{algorithmic}
\usepackage{graphicx}
\usepackage{textcomp}
\usepackage{dsfont}
\usepackage[linesnumberedhidden,ruled,vlined]{algorithm2e}
\SetKwInput{Init}{initialize}
\SetKwInput{KwInput}{{Input}}
\SetKwInOut{Output}{output}
\SetKwRepeat{Repeat}{repeat}{until}

\def\BibTeX{{\rm B\kern-.05em{\sc i\kern-.025em b}\kern-.08em
    T\kern-.1667em\lower.7ex\hbox{E}\kern-.125emX}}

\begin{document}
\title{Accelerated MR Fingerprinting with Low-Rank and Generative Subspace Modeling}
\author{Hengfa Lu, \textit{Student Member}, \textit{IEEE}, Huihui Ye, Lawrence L. Wald, and Bo Zhao, \textit{Member}, \textit{IEEE}
\thanks{This work was supported in part by the National Institutes of Health under Grant NIH-R00-EB027181 and NIH-R01-EB017219.}
\thanks{H. Lu is with the Department of Biomedical Engineering, University of Texas at Austin, Austin, Texas 78712 USA (e-mail: hengfalu@utexas.edu). }
\thanks{H. Ye is with the State Key Laboratory of Modern Optical Instrumentation, College of Optical Science and Engineering, Zhejiang University, Hangzhou, Zhejiang, China, and also with Center for Brain Imaging Science and Technology, Key Laboratory for Biomedical Engineering of Ministry of Education, College of Biomedical Engineering and Instrumental Science, Zhejiang University, Hangzhou, Zhejiang, China (e-mail: yehuihui@zju.edu.cn).}
\thanks{L. L. Wald is with the Athinoula A. Martinos Center for Biomedical Imaging, Massachusetts General Hospital, Charlestown, MA 02129 USA, also with the Department of Radiology, Harvard Medical School, Boston, MA 02115 USA, and also with the Harvard-MIT Division of Health Sciences and Technology, Massachusetts Institute of Technology, Cambridge, MA 02139 USA (e-mail: wald@nmr.mgh.harvard.edu).}
\thanks{B. Zhao is with the Department of Biomedical Engineering and the Oden Institute for Computational Engineering and Sciences, University of Texas at Austin, Austin, Texas 78712 USA (e-mail: bozhao@utexas.edu).} }

\maketitle

\begin{abstract}
Magnetic Resonance (MR) Fingerprinting is an emerging multi-parametric quantitative MR imaging technique, for which image reconstruction methods utilizing low-rank and subspace constraints have achieved state-of-the-art performance. However, this class of methods often suffers from an ill-conditioned model-fitting issue, which degrades the performance as the data acquisition lengths become short and/or the signal-to-noise ratio becomes low. To address this problem, we present a new image reconstruction method for MR Fingerprinting, integrating low-rank and subspace modeling with a deep generative prior. Specifically, the proposed method captures the strong spatiotemporal correlation of contrast-weighted time-series images in MR Fingerprinting via a low-rank factorization. Further, it utilizes an untrained convolutional generative neural network to represent the spatial subspace of the low-rank model, while estimating the temporal subspace of the model from simulated magnetization evolutions generated based on spin physics. Here the architecture of the generative neural network serves as an effective regularizer for the ill-conditioned inverse problem without additional spatial training data that are often expensive to acquire. The proposed formulation results in a non-convex optimization problem, for which we develop an algorithm based on variable splitting and alternating direction method of multipliers. We evaluate the performance of the proposed method with numerical simulations and in vivo experiments and demonstrate that the proposed method outperforms the state-of-the-art low-rank and subspace reconstruction.
\end{abstract}


\section{Introduction}
\label{sec:introduction}
\IEEEPARstart{M}{agnetic} Resonance (MR) Fingerprinting is an emerging quantitative MR imaging technique, which enables the simultaneous quantification of multiple MR tissue parameters (e.g., $T_1$, $T_2$, and proton density) in a single experiment \cite{2013_nature_Ma}. MR Fingerprinting features an innovative transient-state signal excitation strategy. Together with highly-accelerated non-Cartesian acquisitions, it has enabled much faster imaging speeds than conventional quantitative MR imaging techniques. Since the invention of the technique, MR Fingerprinting has found a number of promising applications, including neuroimaging \cite{2018_Radiology_Liao, 2022_Epilepsia_Choi}, cardiovascular imaging \cite{2017_MRM_Hamilton, 2021_JMRI_Jaubert}, cancer imaging \cite{2017_Badve, 2021_Neuroradiol_Pirkl}, and musculoskeletal imaging \cite{2020_MRM_Sharafi}.

The original proof-of-principle MR Fingerprinting reconstruction employs a direct pattern matching scheme \cite{2013_nature_Ma}. Although this method is simple and computationally efficient, it has been shown to be sub-optimal from a statistical estimation perspective \cite{2016_TMI_Zhao}. Various more advanced image reconstruction methods, including model-based \cite{2014_SIAM_IS_Davies,2014_Wang, 2015_ICIP_Zhao, 2016_MRM_Pierre, 2016_TMI_Zhao, 2017_MRI_Doneva, 2018_MRM_Zhao, 2018_MRM_Asslander, 2018_Mazor, 2019_MRM_Lima, 2020_EMBC_Zhao, 2022_TMI_Hu} or learning-based methods \cite{2017_ICIP_Virtue, 2018_MRM_Cohen, 2019_TMI_Fang, 2020_MIA_Balsiger, 2020_NeuroImage_Chen, 2021_MIA_Golbabaee, 2022_EMBC_Lu, 2022_ISBI_Chen, 2023_TMI_Cheng}, have been introduced over the past few years, which have significantly improved the accuracy and/or reduced acquisition times for MR Fingerprinting.

As a class of state-of-the-art image reconstruction methods, the low-rank reconstruction \cite{2015_ICIP_Zhao, 2017_MRI_Doneva, 2018_Mazor, 2018_MRM_Zhao, 2018_MRM_Asslander, 2019_MRM_Lima, 2020_EMBC_Zhao} utilizes a linear low-dimensional model to capture the strong spatiotemporal correlation of contrast-weighted time-series images in MR Fingerprinting. A subspace constraint can be further incorporated into the low-rank model by pre-estimating the temporal subspace of the model from an ensemble of magnetization evolutions simulated based on spin physics \cite{2018_MRM_Zhao, 2018_MRM_Asslander, 2020_EMBC_Zhao}. While the low-rank and subspace reconstruction significantly outperforms the conventional MR Fingerprinting reconstruction, its performance often degrades as the data acquisition length becomes short or the signal-to-noise (SNR) ratio becomes low, due to an ill-conditioned model-fitting issue arising from the undersampling of the temporal subspace of the low-rank model \cite{2012_TMI_Zhao, 2015_MRM_Zhao}.

In this paper, we introduce a new deep learning based method to improve low-rank and subspace reconstruction for MR Fingerprinting. Specifically, the proposed method utilizes an untrained generative convolutional neural network to represent the spatial subspace of the low-rank model, while estimating the temporal subspace directly from an ensemble of magnetization evolutions simulated based on spin physics. Here with the inductive bias of a convolutional neural network architecture, the untrained neural network serves as an effective regularizer for the image reconstruction problem without requiring any additional training data that are often very expensive to collect in MR Fingerprinting. To solve the resulting optimization problem, we develop an algorithm based on variable splitting \cite{2010_TIP_Afonso, 2011_TMI_Ramani} and the alternating direction method of multipliers (ADMM) \cite{2011_FTML_Boyd, 2011_TMI_Ramani}. We evaluate the proposed method with numerical simulations and in vivo experiments and demonstrate the improvement of the proposed method over the state-of-the-art low-rank and subspace reconstruction. A preliminary account of this work was described in our early conference paper \cite{2023_ICASSP_Lu}.

A number of image reconstruction methods have recently been introduced on the use of deep learning to improve low-rank and subspace reconstruction \cite{2020_ISMRM_Sandino, 2022_ISBI_ZihaoChen, 2022_ISMRM_Blumenthal, 2022_TMI_Ahmed}. Note that the proposed method has key differences from these existing methods. First, the proposed method is an unsupervised learning method that utilizes the inductive bias of untrained neural networks as a regularizer. It is different from the supervised learning based reconstruction methods \cite{2020_ISMRM_Sandino, 2022_ISBI_ZihaoChen, 2022_ISMRM_Blumenthal}, which require a relatively large set of training data. Moreover, the proposed method is also distinct from a recent unsupervised learning based reconstruction method \cite{2022_TMI_Ahmed}. Specifically, the proposed method pre-determines the temporal subspace of the low-rank model, which is different from the above method that simultaneously estimates the spatial and temporal subspaces with a deep bi-linear model. In the context of quantitative MRI (e.g., MR Fingerprinting), the temporal subspace is often much easier to determine based on spin physics than dynamic imaging, and a pre-determined temporal subspace results in a simplified image reconstruction problem. In addition, the proposed algorithm is rather different from the one in \cite{2022_TMI_Ahmed}.

The rest of the paper is organized as follows. Section~\ref{sec:proposed_method} describes the proposed method, including the problem formulation and solution algorithm. Section~\ref{sec:results} includes representative results to illustrate the performance of the proposed method. Section~\ref{sec:discussion} contains the discussion, followed by the concluding remarks in Section~\ref{sec:conclusion}.

\section{Proposed Method}
\label{sec:proposed_method}
\subsection{Problem Formulation}
MR Fingerprinting experiments involve acquiring a sequence of contrast-weighted MR images $\{I_{m}(\mathbf{x}) \}_{m=1}^{M}$, in which each image $ I_{m}(\mathbf{x})$ can be parameterized as
\begin{equation}\label{eq:1}
    I_m(\mathbf{x})=\rho(\mathbf{x})\phi_m\left(T_1(\mathbf{x}), T_2(\mathbf{x})\right),
\end{equation}
for $m=1,\cdots,M$, where $\rho(\mathbf{x})$ denotes the proton density, $T_1(\mathbf{x})$ denotes the spin-lattice relaxation time, $T_2(\mathbf{x})$ denotes the spin-spin relaxation time, and $\phi_m (\cdot)$ denotes the contrast-weighting function at the $m$th repetition time (TR), which is determined by the magnetization dynamics governed by the Bloch equation \cite{1946_PhysRev_Bloch}. 

In this work, we consider a discrete image model which assumes that each contrast-weighted image $I_{m}(\mathbf{x})$ can be completely represented by its values at $N$ spatial locations  $\left\{\mathbf{x}_n\right\}_{n = 1}^N$. Accordingly, we can represent the contrast-weighted time-series images $\left\{I_{m}(\mathbf{x})\right\}_{m = 1}^M$ by the following Casorati matrix \cite{2007_ISBI_Liang}:
\begin{equation}\nonumber
\begin{aligned}
    \mathbf{C}
    & =
    \left[\begin{array}{ccc}
    I_1\left(\mathbf{x}_1\right) & \ldots & I_M\left(\mathbf{x}_1\right) \\
    \vdots & \ddots & \vdots \\
    I_1\left(\mathbf{x}_N\right) & \ldots & I_M\left(\mathbf{x}_N\right)
    \end{array}\right] \in \mathbb{C}^{N \times M}.
\end{aligned}
\end{equation}
In MR Fingerprinting, the imaging equation can be written as 
\begin{equation}\label{eq:2}
    \mathbf{d}_{m, c}= \mathbf{F}_{m} \mathbf{S}_c \mathbf{I}_m +\mathbf{n}_{m, c},
\end{equation}
for $m = 1, \cdots, M$ and $c = 1, \cdots, N_c$. Here $\mathbf{I}_m \in \mathbb{C}^N$ denotes the contrast-weighted MR image at the $m$th TR, $\mathbf{F}_m \in \mathbb{C}^{P_m \times N}$ denotes the undersampled Fourier-encoding matrix at the $m$th TR, $\mathbf{S}_c \in \mathbb{C}^{N \times N}$ is a diagonal matrix whose diagonal entries contain the coil sensitivities from the $c$th coil, $\mathbf{d}_{m, c} \in \mathbb{C}^{P_m}$ denotes the measured $\mathbf{k}$-space data from the  $c$th receiver coil at the $m$th TR, and $\mathbf{n}_{m, c} \in \mathbb{C}^{P_m}$ denotes the measurement noise that is assumed to be complex Gaussian noise. 

For notational simplicity, we can rewrite \eqref{eq:2} in a more compact form as follows:
\begin{equation}\label{eq:15}
    \mathbf{d}_c=\Omega \left(\mathbf{F S}_{c} \mathbf{C}\right)+\mathbf{n}_{c},
\end{equation}
where $\mathbf{d}_c \in \mathbb{C}^{P}$ contains all the measured $\mathbf{k}$-space data from the $c$th coil, $\mathbf{F} \in \mathbb{C}^{\tilde{N} \times N}$ denotes the fully-sampled Fourier encoding matrix,  $\Omega(\cdot): \mathbb{C}^{\tilde{N}\times M} \rightarrow \mathbb{C}^P$ denotes the undersampling operator that acquires the $\mathbf{k}$-space data for each contrast-weighted image and then concatenates them into the data vector $\mathbf{d}_c$, $\mathbf{n}_c \in \mathbb{C}^P$ contains the measurement noise from the $c$th coil, and $P = \sum_{m = 1}^M P_m$.

In MR Fingerprinting experiments, $\mathbf{k}$-space data are highly-undersampled, i.e., $P \ll NM$. A significant technical challenge arises from reconstructing contrast-weighted time-series images from highly-undersampled data. To ensure high-quality reconstruction, we impose a low-rank constraint on $\mathbf{C}$, i.e., $\mathrm{rank}(\mathbf{C}) \leq L \ll \mathrm{min}\left\{N, M\right\}$, by exploiting the strong spatiotemporal correlation of contrast-weighted images in MR Fingerprinting. The low-rank constraint reduces the number of degrees of freedom in $\mathbf{C}$ to $L(M+N-L)$ \cite{2010_SIAM_Recht}, which is often much less than the total number of entries of $\mathbf{C}$. This enables image reconstruction from highly-undersampled $\mathbf{k}$-space data. 

While there are many ways of imposing a low-rank constraint \cite{2010_SIAM_Recht}, here we use a low-rank matrix factorization \cite{2010_ISBI_Zhao, 2010_ISBI_Haldar, 2012_TMI_Zhao}, i.e., $\mathbf{C} = \mathbf{U}\mathbf{V}$, where $\mathbf{U} \in \mathbb{C}^{N \times L}$ and $\mathbf{V} \in \mathbb{C}^{L \times M}$. Note that the columns of $\mathbf{U}$ and the rows of $\mathbf{V}$ respectively span the spatial subspace and the temporal subspace of $\mathbf{C}$. Further, we can incorporate a subspace constraint by pre-estimating $\mathbf{V}$ from an ensemble of magnetization evolutions simulated based on spin physics \cite{2018_MRM_Zhao} or some physically-acquired navigator data \cite{2023_ISMRM_Lu}. 

\begin{figure}[!t]
\centerline{\includegraphics[width=\columnwidth]{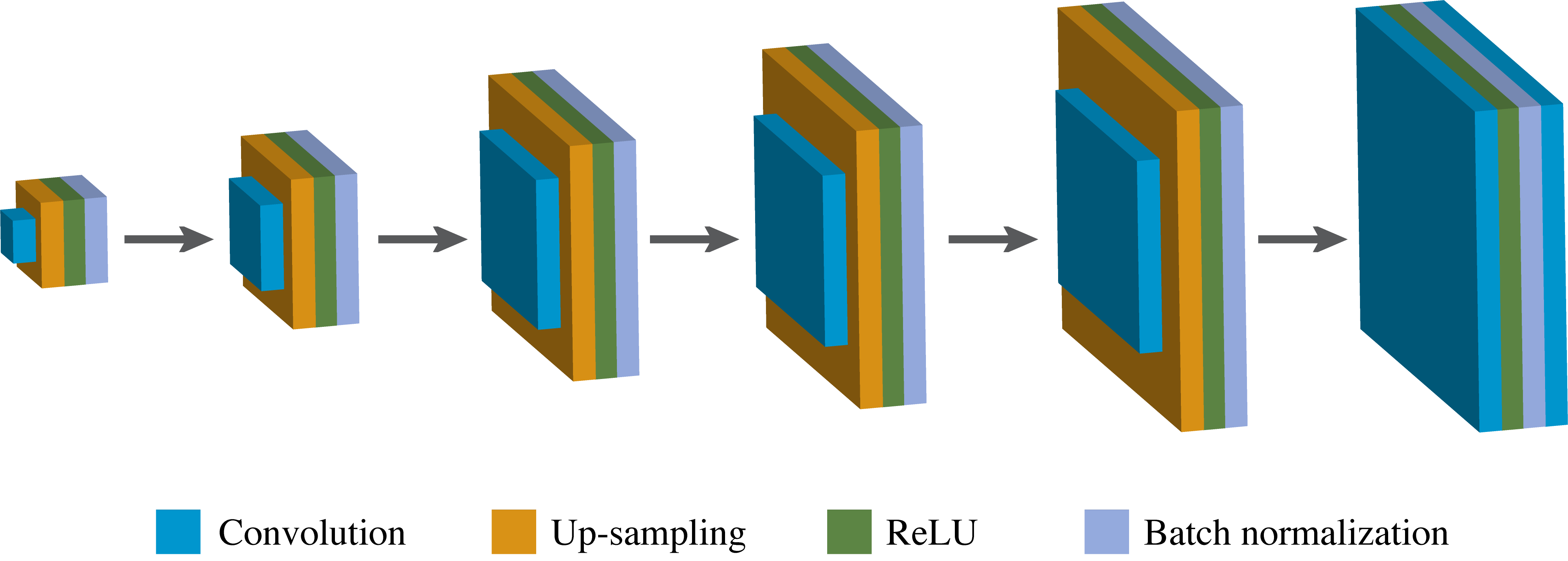}}
\caption{Architecture of a generative neural network in the proposed method, which consists of convolutional, up-sampling, ReLu, and batch normalization layers.}
\label{fig:network_architecture}
\end{figure}
While the low-rank and subspace reconstruction significantly outperforms the dictionary-matching based conventional reconstruction \cite{2013_nature_Ma}, it often suffers from an ill-conditioned model fitting issue, due to the undersampling of the temporal subspace of the model. This degrades the performance of the low-rank and subspace reconstruction \cite{2018_MRM_Zhao}, as the data acquisition length becomes short and/or SNR becomes low.

To address the issue, here we incorporate a deep generative prior into the low-rank and subspace reconstruction. While the effectiveness of a pre-trained deep generative prior \cite{2017_PMLR_Bora} has been demonstrated in solving inverse problems, it often requires large training data sets, which are often very expensive to acquire in MR Fingerprinting. Recently, it has been demonstrated that with the inductive bias of a deep convolutional architecture, an untrained generative neural networks (e.g., deep image prior \cite{2018_CVPR_Ulyanov} and deep decoder \cite{2018_ICLR_Heckel,2021_TCI_Darestani}) can serve as an effective spatial prior. Here we use an untrained generative neural network to represent the spatial subspace of a low-rank model, which serves as a regularizer for the low-rank and subspace reconstruction. Mathematically, we have $\mathbf{U} = \mathbf{G}_{\boldsymbol{\theta}}(\mathbf{z})$, where $\mathbf{G}_{\boldsymbol{\theta}}(\mathbf{z}): \mathbb{R}^{N_{0}} \rightarrow \mathbb{R}^{N \times L}$ is a generative neural network, $\mathbf{z} \in \mathbb{R}^{N_{0}}$ is a low-dimensional latent random vector, and $\boldsymbol{\theta}$ contains the trainable parameters of the neural network. Here we illustrate an example generative convolutional neural network architecture in Fig. 1, which consists of convolutional, up-sampling, ReLu, and batch normalization layers.

Incorporating the low-rank and subspace model with the deep generative prior, the image reconstruction problem can be formulated as follows:
\begin{equation} \label{eq:proposed_formula}
\begin{aligned}
   \left \{ \hat{\mathbf{U}}, \hat{\boldsymbol{\theta}} \right \}  & = \arg \min_{\mathbf{U}, \boldsymbol{\theta}}  \sum_{c=1}^{N_{c}} \left\|\mathbf{d}_{c}-\Omega \left(\mathbf{F} \mathbf{S}_{c} \mathbf{U} \hat{\mathbf{V}} \right)\right\|_2^2, \\
   & \ \qquad \textrm{s.t.} \quad \mathbf{U} = \mathbf{G}_{\boldsymbol{\theta}}(\mathbf{z}),
\end{aligned}
\end{equation}
where $\hat{\mathbf{V}}$ denotes the pre-estimated temporal subspace of the low-rank model by singular value decomposition. After solving \eqref{eq:proposed_formula}, we can reconstruct the contrast-weighted time-series images by $\hat{\mathbf{I}} = \hat{\mathbf{U}} \hat{\mathbf{V}}$, from which we then estimate MR tissue parameters via a voxel-wise pattern matching. Next, we describe an optimization algorithm to solve \eqref{eq:proposed_formula}.

\subsection{Solution Algorithm}
The proposed formulation in \eqref{eq:proposed_formula} results in a non-convex optimization problem. Here we describe an algorithm based on variable splitting and the ADMM method \cite{2011_FTML_Boyd, 2011_TMI_Ramani} to solve this problem. First, we introduce a set of auxiliary variables $\left\{\mathbf{H}_{c}\right\}_{c=1}^{N_{c}}$ to split the coil sensitivities from the objective function. This converts \eqref{eq:proposed_formula} into the following equivalent constrained optimization problem:
\begin{equation} \label{eq:proposed_formula_split}
\begin{aligned}
    & \min_{\mathbf{U}, \left \{\mathbf{H}_{c}\right\}, \boldsymbol{\theta} }  ~\sum_{c=1}^{N_{c}} \left\|\mathbf{d}_{c}-\Omega \left(\mathbf{F} \mathbf{H}_{c} \right)\right\|_2^2, \\
    & \ \textrm{s.t.} \quad \mathbf{H}_{c} = \mathbf{S}_{c}\mathbf{U}\hat{\mathbf{V}} \quad \textrm{and} \quad \mathbf{U} = \mathbf{G}_{\boldsymbol{\theta}}(\mathbf{z}),
\end{aligned}
\end{equation}
for $c=1, \cdots, N_{c}$.

Second, we form the augmented Lagrangian associated with \eqref{eq:proposed_formula_split} as follows:
\begin{equation} \label{eq:AL}
\begin{aligned}
     & \mathcal{L}\left ( \left\{\mathbf{H}_{c}\right\}, \mathbf{U}, \boldsymbol{\theta}, \left\{\mathbf{\Lambda}_{c} \right\}, \mathbf{\Gamma} \right )
     = \sum_{c=1}^{N_{c}} \Big \{ \left\|\mathbf{d}_{c}-\Omega \left(\mathbf{F} \mathbf{H}_{c} \right)\right\|_2^{2} + \\
     & \textbf{Re} ( < \mathbf{\Lambda}_{c}, \mathbf{H}_{c} - \mathbf{S}_{c}\mathbf{U}\hat{\mathbf{V}} > ) + \frac{\mu_{1}}{2} \left \| \mathbf{H}_{c} - \mathbf{S}_{c}\mathbf{U}\hat{\mathbf{V}} \right \|_{F}^{2} \Big \} +  \\
     & \textbf{Re} ( < \mathbf{\Gamma}, \mathbf{U} - \mathbf{G}_{\boldsymbol{\theta}}(\mathbf{z}) > ) + \frac{\mu_{2}}{2} \left \| \mathbf{U} - \mathbf{G}_{\boldsymbol{\theta}}(\mathbf{z}) \right \|_{F}^{2},
\end{aligned}
\end{equation}
where $\left\{ \mathbf{\Lambda}_{c} \right\}_{c=1}^{N_{c}}$ and $\mathbf{\Gamma}$ are the Lagrangian multipliers, $\mu_{1}, \mu_{2}>0$ are the penalty parameters, and the operator $\textbf{Re} ( \cdot )$ takes the real part of a complex number.

Next, we iteratively minimize \eqref{eq:AL} by solving the following three subproblems:
\begin{equation}\label{eq:subproblem1}
    \begin{aligned}
    \mathbf{H}_{c}^{(k+1)} = \arg \min_{\left\{\!\mathbf{H}_{c} \!\right\}} \mathcal{L} \! \left (  \left\{\mathbf{H}_{c}\right\}, \mathbf{U}^{(k)}, \boldsymbol{\theta}^{(k)}, \{\! \mathbf{\Lambda}_{c}^{(k)}\}, \mathbf{\Gamma}^{(k)} \! \right ),
    \end{aligned}
\end{equation}
\begin{equation} \label{eq:subproblem2}
    \begin{aligned}
    \mathbf{U}^{(k+1)} = \arg \min_{\mathbf{U}} \mathcal{L} \! \left ( \! \{ \mathbf{H}_{c}^{(k+1)}  \}, \mathbf{U}, \boldsymbol{\theta}^{(k)}, \{ \mathbf{\Lambda}_{c}^{(k)} \}, \mathbf{\Gamma}^{(k)} \! \right ),
    \end{aligned}
\end{equation}
\begin{equation} \label{eq:subproblem3}
    \begin{aligned}
    \boldsymbol{\theta}^{(k+1)} = \arg \min_{\boldsymbol{\theta}} \mathcal{L} \! \left ( \! \{ \! \mathbf{H}_{c}^{(k+1)} \!\}, \mathbf{U}^{(k+1)}, \boldsymbol{\theta}, \{ \! \mathbf{\Lambda}_{c}^{(k)} \! \}, \mathbf{\Gamma}^{(k)} \! \right ),
    \end{aligned}
\end{equation}
and updating the Lagrangian multipliers as follows:
\begin{equation}\label{eq:multiplier1}
    \begin{aligned}
    \mathbf{\Lambda}_{c}^{(k+1)} = \mathbf{\Lambda}_{c}^{(k)} + \mu_{1} \left( \mathbf{H}_{c}^{(k+1)} - \mathbf{S}_{c}\mathbf{U}^{(k+1)}\hat{\mathbf{V}} \right),
    \end{aligned}
\end{equation}
\begin{equation}\label{eq:multiplier2}
    \begin{aligned}
    \mathbf{\Gamma}^{(k+1)} = \mathbf{\Gamma}^{(k)} + \mu_{2} \left( \mathbf{U}^{(k+1)} - \mathbf{G}_{\boldsymbol{\theta}^{(k+1)}}(\mathbf{z}) \right),
    \end{aligned}
\end{equation}
until the change of the solution is less than a pre-specified tolerance or the maximum number of iterations is reached. Next, we describe the detailed solutions to the subproblems in \eqref{eq:subproblem1} - \eqref{eq:subproblem3}.

\textit{1) Solution to \eqref{eq:subproblem1}}: Note that the subproblem in \eqref{eq:subproblem1} can be written as
\begin{equation} \label{eq:subproblem1-1}
     \min_{\{\mathbf{H}_{c}\} } \sum_{c=1}^{N_{c}} \left \{ \left\|\mathbf{d}_{c}-\Omega \left(\mathbf{F} \mathbf{H}_{c} \right)\right\|_2^{2} + \frac{\mu_{1}}{2} \left \| \mathbf{H}_{c} - \mathbf{Q}_{c}^{(k)}\right \|_{F}^{2} \right \},
\end{equation}
where $\mathbf{Q}_{c}^{(k)} = \mathbf{S}_{c}\mathbf{U}^{(k)}\hat{\mathbf{V}} - \mathbf{\Lambda}_{c}^{(k)}/{\mu_{1}}$. This is a linear least-squares problem separable for each $\mathbf{H}_c$, i.e.,
\begin{equation} \label{eq:subproblem1-2}
     \min_{\mathbf{H}_{c} } \left\|\mathbf{d}_{c}-\Omega \left(\mathbf{F} \mathbf{H}_{c} \right)\right\|_2^{2} + \frac{\mu_{1}}{2} \left \| \mathbf{H}_{c} - \mathbf{Q}_{c}^{(k)}\right \|_{F}^{2},
\end{equation}
for $c=1, \cdots, N_{c}$. Further, it can be shown that \eqref{eq:subproblem1-2} is separable for each column of $\mathbf{H}_c$, i.e., 
\begin{equation} \label{eq:ls}
     \min_{\mathbf{h}_{m, c} } \left\|\mathbf{d}_{m, c}- \mathbf{F}_m \mathbf{h}_{m, c} \right\|_2^{2} + \frac{\mu_{1}}{2} \left \| \mathbf{h}_{m, c} - \mathbf{q}_{m, c}^{(k)}\right \|_{2}^{2},
\end{equation}
for $m = 1, \cdots, M$, where $\mathbf{h}_{m, c}$ and $\mathbf{q}^{(k)}_{m, c}$ are the $m$th columns of $\mathbf{H}_c$ and $\mathbf{Q}^{(k)}_c$, respectively. The solution to \eqref{eq:ls} can be written as 
\begin{equation} \label{eq:subproblem1-3}
    \left ( \mathbf{F}_m^{H} \mathbf{F}_m + \frac{\mu_1}{2}\mathds{1}_{N} \right ) \mathbf{h}_{m, c} = \mathbf{F}_m^{H}\mathbf{d}_{m, c} + \frac{\mu_{1}}{2} \mathbf{q}_{m, c}^{(k)},
\end{equation}
where $\mathds{1}_{N} \in \mathbb{C}^{N \times N}$ is the identity matrix. While \eqref{eq:subproblem1-3} can be solved iteratively (e.g., with the conjugate gradient algorithm \cite{2006_book_nocedal}), we can solve it in a more efficient way with a direct matrix inversion, given that the same coefficient matrix $(\mathbf{F}_m^H\mathbf{F}_m + \mu_1\mathds{1}_{N}/2)$ is shared by a number of linear systems of equations at different iterations of the algorithm\footnote{In standard MR Fingerprinting acquisitions (e.g., \cite{2013_nature_Ma, 2015_MRM_Jiang}), the same $\mathbf{k}$-space trajectory repeats at different TRs. Thus, the same coefficient matrix $(\mathbf{F}_m^H\mathbf{F}_m + \mu_1\mathds{1}_{N}/2)$ also occurs at different $m$ in each iteration of the algorithm.}. 

Specifically, consider the inversion of the coefficient matrix of \eqref{eq:subproblem1-3}, and, by applying the matrix inversion lemma \cite{2005_book_bernstein}, we have 
\begin{equation}
 \begin{aligned} \label{eq:subproblem1-4}
     & \left ( \mathbf{F}_{m}^{H} \mathbf{F}_{m} + \frac{\mu_{1}}{2}\mathds{1}_N \right )^{-1} \\
     & = \frac{2}{\mu_{1}} \mathds{1}_N - \mathbf{F}_{m}^{H} \left [ \left( \frac{\mu_{1}}{2} \right)^{2} \mathds{1}_N + \frac{\mu_{1}}{2} \mathbf{F}_{m} \mathbf{F}_{m}^{H} \right ]^{-1} \mathbf{F}_{m}.
 \end{aligned}
 \end{equation}
Note that $\mathbf{F}_m\mathbf{F}^H_m \in \mathbb{C}^{P_m \times P_m}$ is a  small-scale matrix that can be pre-computed and saved, given that $\mathbf{k}$-space data are highly-undersampled for each $\mathbf{I}_m$ (i.e., $P_m \ll \tilde{N}$). Here we only need to compute the matrix $\left [ \left( \mu_{1} / 2 \right)^{2} \mathds{1}_N + \mu_{1} / 2 \mathbf{F}_{m} \mathbf{F}_{m}^{H} \right ]^{-1}$ once and save it for later use at different iterations of the algorithm. Consequently, the matrix inversion in \eqref{eq:subproblem1-4} only encompasses two non-uniform Fourier transforms and some small-scale matrix-vector multiplications that can be calculated very efficiently.

\textit{2) Solution to \eqref{eq:subproblem2}}: The subproblem in \eqref{eq:subproblem2} can be written as
\begin{equation} \label{eq:subproblem2-1}
\begin{aligned}
    \min_{\mathbf{U}} & \sum_{c=1}^{N_{c}} \left \| \mathbf{S}_{c}\mathbf{U}\hat{\mathbf{V}}  - \left( \mathbf{H}_{c}^{(k+1)} + 1/\mu_{1} \mathbf{\Lambda}_{c}^{(k)} \right)\right \|_{F}^{2} + \\
    & \qquad \qquad \qquad \qquad \qquad \quad \frac{\mu_{2}}{\mu_{1}} \left \| \mathbf{U} - \mathbf{G}_{\boldsymbol{\theta}^{(k)}}(\mathbf{z}) \right \|_{F}^{2},
\end{aligned}
\end{equation}
which can be solved by 
\begin{equation} \label{eq:subproblem2-3}
    \left(  \sum_{c=1}^{N_c} {\mathbf{S}}_c^H {\mathbf{S}}_c + \frac{\mu_{2}}{\mu_1} \mathds{1}_N \right) \mathbf{U} = \mathbf{W}^{(k+1)},
\end{equation}
where
\begin{equation}\nonumber
    \mathbf{W}^{(k+1)} = \sum_{c=1}^{N_{c}} {\mathbf{S}}_c^H \left( \mathbf{H}_{c}^{(k+1)} + \frac{\mathbf{\Lambda}_{c}^{(k)}}{\mu_{1}} \right)\hat{\mathbf{V}}^{H} + \frac{\mu_{2}}{\mu_{1}}\mathbf{G}_{\boldsymbol{\theta}^{(k)}}(\mathbf{z}).
\end{equation}
Noting that the coefficient matrix $\left(  \sum_{c=1}^{N_c} {\mathbf{S}}_c^H {\mathbf{S}}_c + \mu_{2}/\mu_1 \mathds{1}_N \right)$
is diagonal, we can solve \eqref{eq:subproblem2-3} very efficiently via an entry-by-entry inversion.

\textit{3) Solution to \eqref{eq:subproblem3}}: The subproblem in \eqref{eq:subproblem3} can be written as
\begin{equation} \label{eq:subproblem3-1}
     \min_{\boldsymbol{\theta}} \left \|  \mathbf{G}_{\boldsymbol{\theta}}(\mathbf{z}) - \left( \mathbf{U}^{(k+1)} + \frac{1}{\mu_{2}} \mathbf{\Gamma}^{(k)}\right ) \right \|_{F}^{2}.
\end{equation}
This is a large-scale nonlinear optimization problem similar to a neural network training problem. Here we solve it by the Adam algorithm \cite{2015_ICLR_Kingma}.

For clarity, we summarize the overall algorithm in Algorithm~\ref{alg:proposed}. Given that \eqref{eq:proposed_formula_split} is a non-convex optimization problem, the performance of the proposed algorithm depends on initialization. Here we initialize $\mathbf{U}$ using the low-rank and subspace reconstruction, which is often computationally efficient. Additionally, we initialize the Lagrangian multipliers $\mathbf{\Lambda}_{c}$ and  $\mathbf{\Gamma}$ with zero matrices and vector, and initialize $\boldsymbol{\theta}$ with a random vector. In the next section, we will illustrate the performance of the above initialization scheme.

\begin{algorithm}\label{alg:proposed}
\DontPrintSemicolon
\KwInput{Measured data $\{\mathbf{d}_{c}\}$, sampling operator $\Omega$, coil sensitivities $\left \{ \mathbf{S}_{c} \right \}$, temporal subspace $\hat{\mathbf{V}}$, neural network $\mathbf{G}_{\boldsymbol{\theta}}(\mathbf{z})$, and penalty parameters $\mu_{1}$, $\mu_{2}$.} \vspace{6pt}
\Init{$\mathbf{U}^{(0)}$, $\{ \mathbf{\Lambda}^{(0)}_{c} \}$, $\mathbf{\Gamma}^{(0)}$, and $\boldsymbol{\theta}^{(0)}$.}
\Repeat{convergence}{
       a. Update $ \mathbf{H}_{c}^{(k+1)}$ via \eqref{eq:subproblem1-3}. \\ \vspace{4pt}
       b. Update $\mathbf{U}^{(k+1)}$ via \eqref{eq:subproblem2-3}. \\ \vspace{4pt}
       c. Update $\boldsymbol{\theta}^{(k+1)}$ via \eqref{eq:subproblem3-1}. \\ \vspace{4pt}
       d. \vspace{-3.0mm}
       \begin{equation}\nonumber
            \begin{aligned}
            \mathbf{\Lambda}_{c}^{(k+1)} & = \mathbf{\Lambda}_{c}^{(k)} + \mu_{1} \left( \mathbf{H}_{c}^{(k+1)} - \mathbf{S}_{c}\mathbf{U}^{(k+1)}\hat{\mathbf{V}} \right), \\
            \mathbf{\Gamma}^{(k+1)} & = \mathbf{\Gamma}^{(k)} + \mu_{2} \left( \mathbf{U}^{(k+1)} - \mathbf{G}_{\boldsymbol{\theta}^{(k+1)}}(\mathbf{z}) \right).
            \end{aligned}
        \end{equation}
   }
\vspace{4pt}
\Output{$\hat{\mathbf{U}}$}
\caption{Proposed algorithm for \eqref{eq:proposed_formula}}
\end{algorithm}


\section{Results}
\label{sec:results}
In this section, we show representative results from simulations and in vivo experiments to demonstrate the performance of the proposed method.

\subsection{Simulation Results}
We conducted simulations on a numerical brain phantom to evaluate the proposed method. The numerical phantom simulates a single-channel MR Fingerprinting experiment, which has the same setup as the one in \cite{2016_TMI_Zhao}. Specifically, we took the $T_1$, $T_2$, and proton density maps from the Brainweb database \cite{1998_TMI_Collins} as the ground truth. We performed Bloch simulations to generate contrast-weighted time-series images by an inversion recovery fast imaging with steady state precession (IR-FISP) sequence \cite{2015_MRM_Jiang}. Here we used the same set of flip angles and repetition times as in \cite{2015_MRM_Jiang} for Bloch simulations. We set the field-of-view (FOV) as $300 \times 300$ mm$^2$ and the corresponding matrix size as $256 \times 256$. 

We acquired highly-undersampled $\mathbf{k}$-space data using one spiral interleaf per TR and a set of fully-sampled $\mathbf{k}$-space data consists of $48$ spiral interleaves. We added complex white Gaussian noise to measured data. Here we define the signal-to-noise ratio (SNR) as $\mathrm{SNR} = 20 \log_{10}(s/\sigma)$, where $s$ denotes the average signal intensity within a region of the white matter in the first contrast-weighted image, and $\sigma$ denotes the standard deviation of the noise.

We performed image reconstruction using the low-rank and subspace method \cite{2018_MRM_Zhao} and the proposed method. Here we manually selected the rank $L=6$ for both methods for optimized performance. For the proposed method, we employed a generative convolutional neural network architecture illustrated in Fig.~\ref{fig:network_architecture}. We solved the subproblem \eqref{eq:subproblem3-1} using the Adam algorithm with the learning rate $ = 0.01$ and the number of iterations $ = 300$. In addition, we manually selected the penalty parameters of the ADMM algorithm for optimized performance. We implemented both reconstruction methods on a Linux server with dual Intel Xeon Platinum $8362$R processors (each with $2.80$ GHz and $32$ cores), $2.00$ TB RAM, and dual NVIDIA A$100$ GPUs (each with $80$ GB RAM) running MATLAB$^\circledR$ R2022b and Deep Learning Toolbox\textsuperscript{TM}.

After image reconstruction, we estimated MR tissue parameter maps by dictionary matching. Here the same dictionary was used for both methods based on the following discretization scheme: 1) the range of possible $T_{1}$ values was in $[100, 3000]$ ms, with an incremental step of $10$ ms in the range of $[100, 1500]$ ms and an incremental step of $20$ ms in the range of $[1501, 3000]$ ms; and 2) the range of possible $T_{2}$ values was in $[20, 350]$ ms, with an incremental step of $1$ ms in the range of $[20, 200]$ ms and an incremental step of $2$ ms in the range of $[201, 350]$ ms.

To assess the accuracy of reconstructed contrast-weighted MR images or parameter maps, we used the normalized root-mean-square error (NRMSE) defined as $\| \mathbf{g} - \hat{\mathbf{g}} \|_{2} / \| \mathbf{g} \|_{2}$,\footnote{The regions associated with the background, skull, scalp, and cerebrospinal fluid were not included into the NRMSE calculation, given that they are often not regions of interest for neuroimaging studies.} where $\mathbf{g}$ and $\hat{\mathbf{g}}$ respectively denote the ground truth and the reconstruction. 

\begin{figure}[!t]
\centerline{\includegraphics[width=0.7\columnwidth]{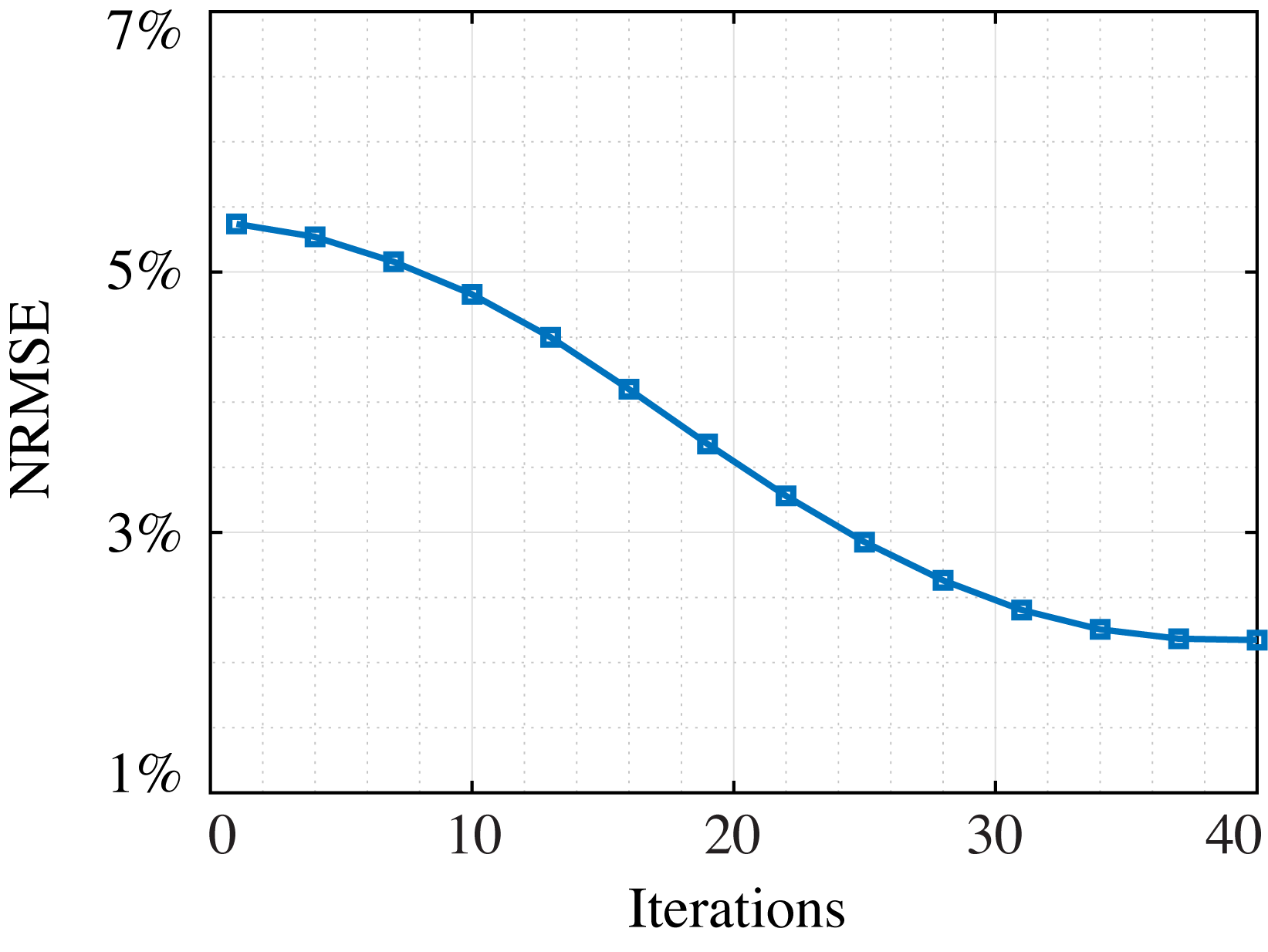}}
\caption{The NRMSE of reconstructed time-series images at different iterations of the proposed algorithm at the acquisition length $M = 400$ and $\mathrm{SNR} = 33~$dB.}
\label{fig:diff_init}
\end{figure}

\begin{figure}[!t]
\centerline{\includegraphics[width=\columnwidth]{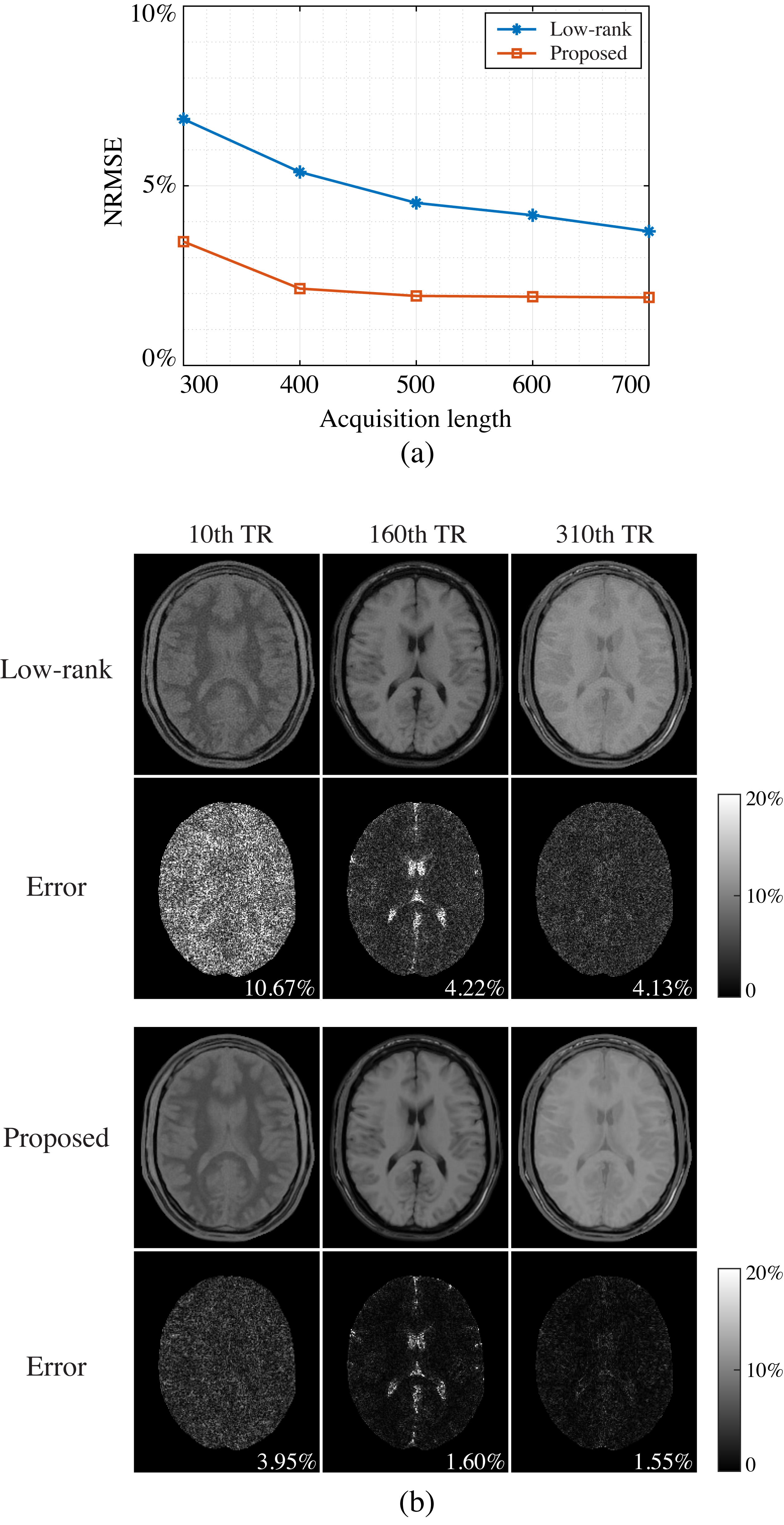}}
\caption{(a) The NRMSE of reconstructed MR time-series images using the low-rank and subspace reconstruction and the proposed method at different acquisition lengths. (b) Reconstructed contrast-weighted images for three representative TRs and the associated relative error maps at the acquisition length $M = 400$ and $\mathrm{SNR} = 33~$dB. Note that the NRMSE is labeled at the right corner of each error map.}
\label{fig:TS_image_nrmse_acqs}
\end{figure}
\begin{figure*}[!t]
\centerline{\includegraphics[width=\textwidth]{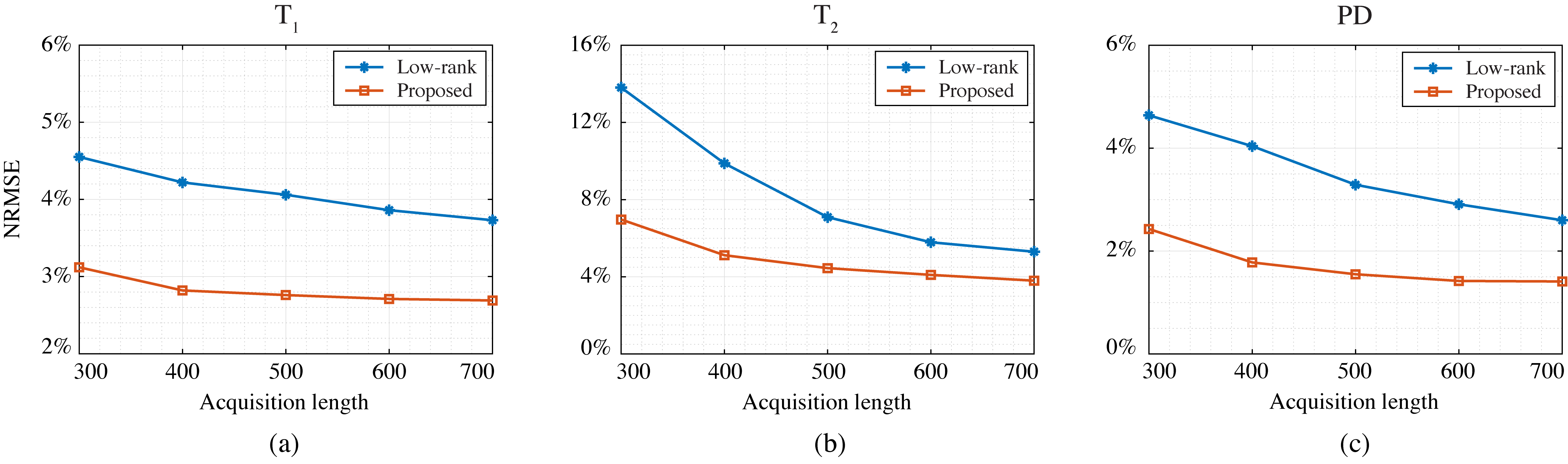}}
\caption{The NRMSE for the reconstruction of (a) $T_1$, (b) $T_2$, and (c) proton density maps from the low-rank and subspace reconstruction and the proposed method at different acquisition lengths at $\mathrm{SNR} = 33~$dB.}
\label{fig:nrmse_acq_lens}
\end{figure*}

\begin{figure*}[!t]
\centerline{\includegraphics[width=0.9\textwidth]{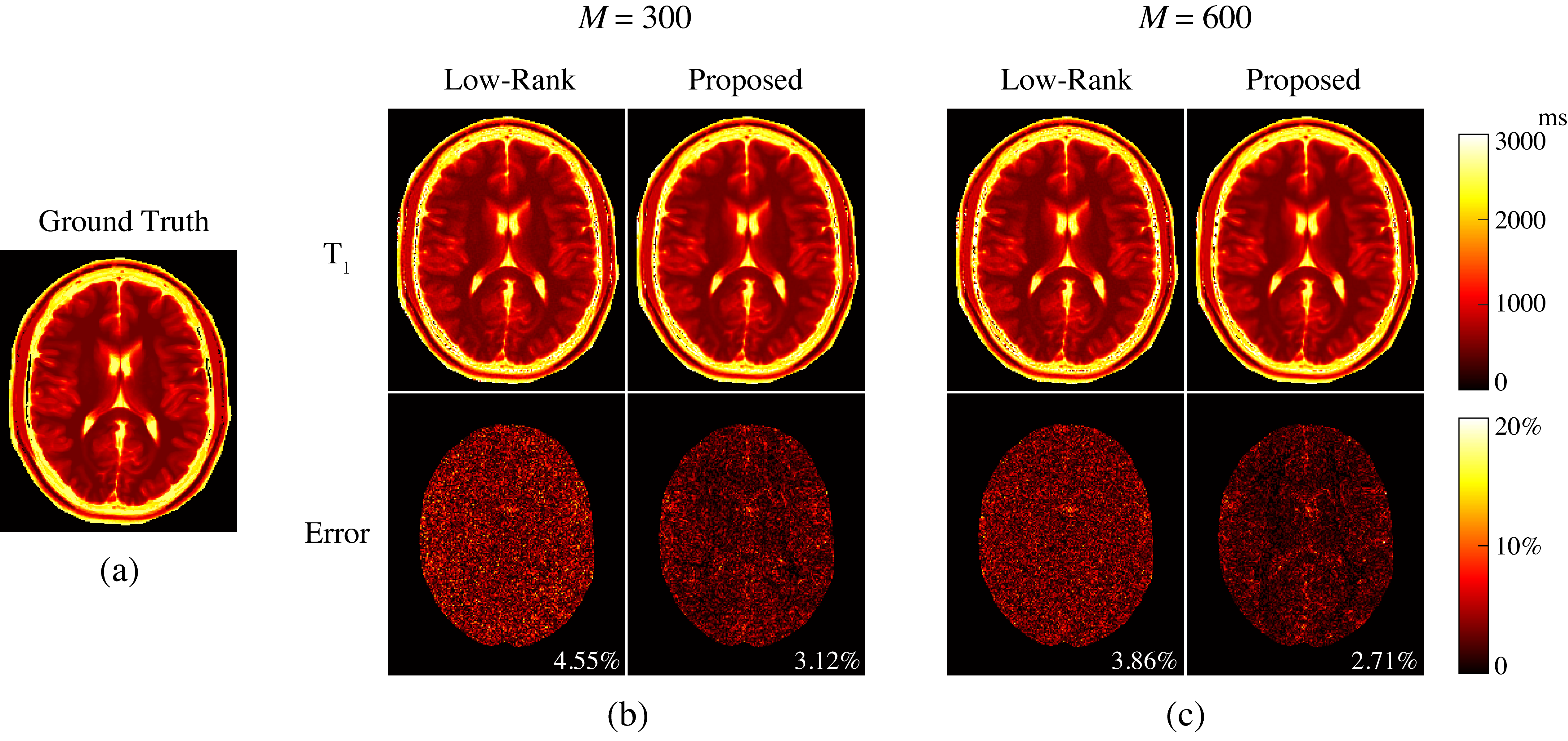}}
\caption{Reconstructed $T_{1}$ maps from the low-rank and subspace reconstruction and the proposed method with two acquisition lengths, i.e., $M=300$ and $600$. (a) Ground-truth $T_{1}$ map. (b) Reconstructed $T_{1}$ maps and corresponding error maps with $M=300$. (c) Reconstructed $T_{1}$ maps and corresponding error maps with $M=600$. Note that the NRMSE is labeled at the right corner of each error map.}
\label{fig:recon_T1_map_phantom}
\end{figure*}

\subsubsection{Algorithm convergence}
The proposed formulation results in a non-convex optimization problem. While there is no theoretical guarantee for the convergence of the algorithm for solving \eqref{eq:proposed_formula}, we demonstrate the convergence empirically with numerical simulations. Specifically, we simulated an MR Fingerprinting experiment with an acquisition length $M = 400$ and $\mathrm{SNR} = 33~$dB, and performed image reconstruction using the proposed method. Here we initialized $\mathbf{U}$ using the low-rank and subspace reconstruction. We initialized $\mathbf{\Lambda}_{c}$ and  $\mathbf{\Gamma}$ with zero matrices and vector and $\boldsymbol{\theta}$ with a random vector. Fig.~\ref{fig:diff_init} shows the NRMSE of the reconstructed time-series images using the proposed method. As can be seen, with a relatively small number of iterations, the proposed method improves the NRMSE by a factor of two, as compared to the initialization with the low-rank and subspace reconstruction. 

\subsubsection{Impact of acquisition length}
We evaluated the performance of the low-rank and subspace reconstruction and the proposed method at different acquisition lengths, ranging from $M = 300$ to $700$. We set the $\mathrm{SNR} = 33~$dB for this set of simulations. Fig.~\ref{fig:TS_image_nrmse_acqs}(a) shows the NRMSE of the reconstructed time-series images at different acquisition lengths using the above two methods. Fig.~\ref{fig:TS_image_nrmse_acqs}(b) shows three representative contrast-weighted MR images reconstructed by the above methods. As can be seen, the proposed method substantially outperforms the low-rank and subspace method, which illustrates the benefits of incorporating a deep generative neural network as a regularizer for the low-rank and subspace reconstruction. 

Fig.~\ref{fig:nrmse_acq_lens} shows the NRMSE of the reconstructed $T_1$, $T_2$, and proton density maps from the low-rank and subspace reconstruction and the proposed method at different acquisition lengths. As can be seen, the proposed method improves over the low-rank and subspace method for the reconstruction of $T_1$, $T_2$, proton density maps at all acquisition lengths, and the improvement is more significant for $T_2$, as the acquisition length becomes short. 

\begin{figure*}[!t]
\centerline{\includegraphics[width=0.9\textwidth]{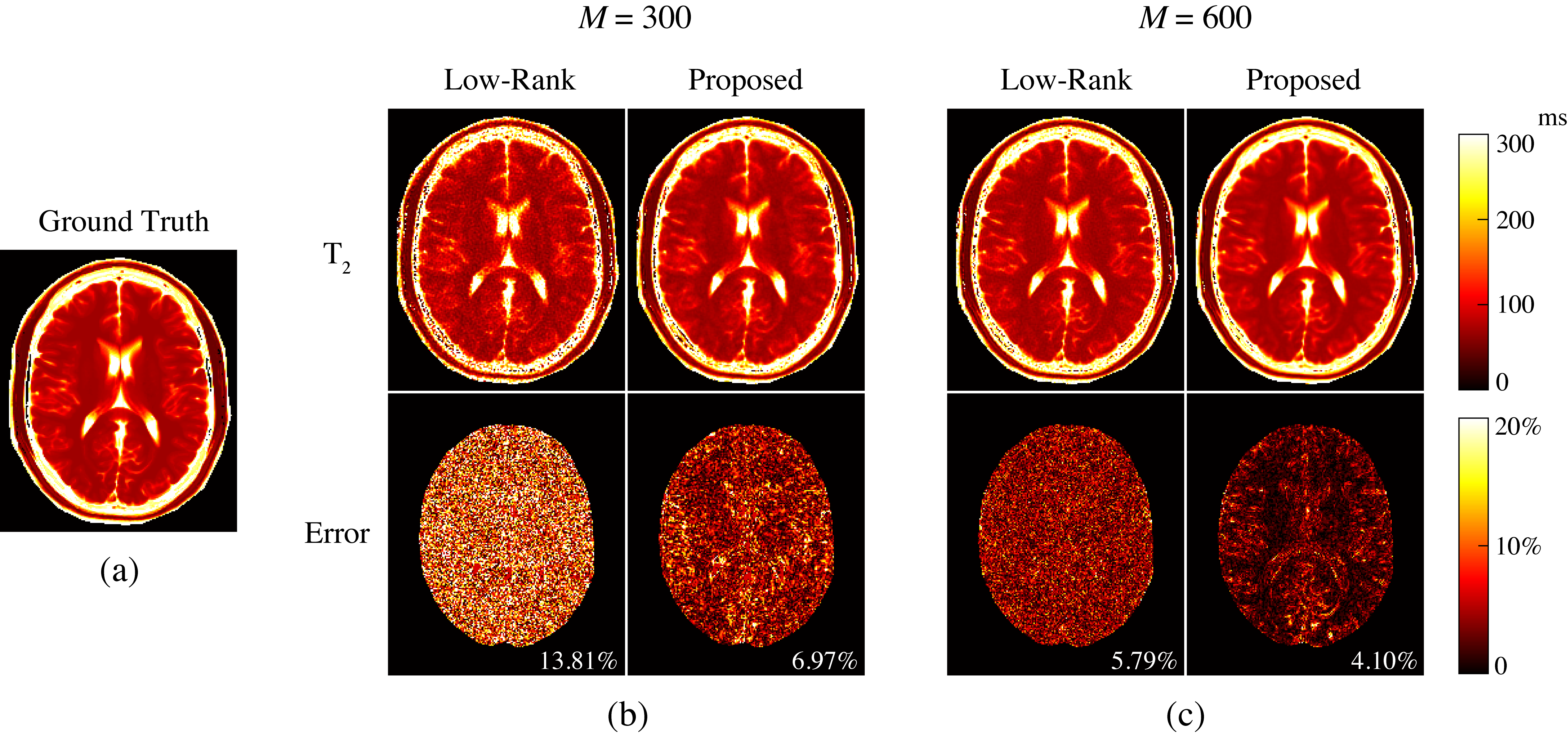}}
\caption{Reconstructed $T_2$ maps from the low-rank and subspace reconstruction and the proposed method with two acquisition lengths, i.e., $M=300$ and $600$. (a) Ground-truth $T_{2}$ map. (b) Reconstructed $T_{2}$ maps and corresponding error maps with $M=300$. (c) Reconstructed $T_{2}$ maps and corresponding error maps with $M=600$. Note that the NRMSE is labeled at the right corner of each error map.}
\label{fig:recon_T2_map_phantom}
\end{figure*}

\begin{figure*}[!t]
\centerline{\includegraphics[width=0.9\textwidth]{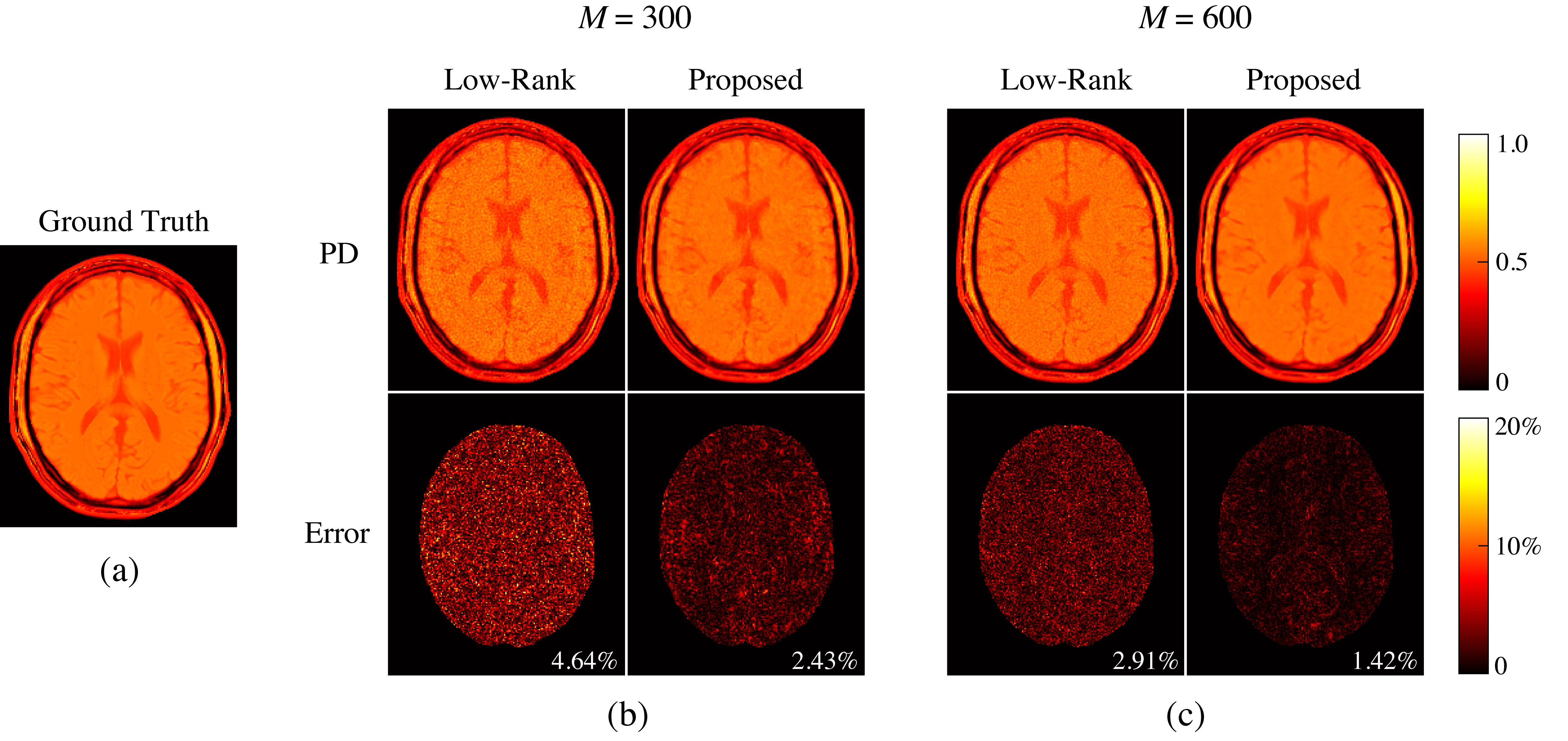}}
\caption{Reconstructed proton density maps from the low-rank and subspace reconstruction and the proposed method with two acquisition lengths, i.e., $M=300$ and $600$. (a) Ground-truth proton density map. (b) Reconstructed proton density maps and corresponding error maps with $M=300$. (c) Reconstructed proton density maps and corresponding error maps with $M=600$. Note that the NRMSE is labeled at the right corner of each error map.}
\label{fig:recon_PD_map_phantom}
\end{figure*}

Figs.~\ref{fig:recon_T1_map_phantom}, \ref{fig:recon_T2_map_phantom}, and \ref{fig:recon_PD_map_phantom} show the reconstructed $T_1$, $T_2$, and proton density maps from the low-rank and subspace reconstruction and the proposed method at the two acquisition lengths (i.e., $M = 300$ and $600$). As can be seen, the proposed method outperforms the low-rank and subspace reconstruction at both acquisition lengths. At a short acquisition length (i.e., $M = 300$), it nicely enables a factor of two improvement for the reconstruction of $T_2$ and proton density maps. It is clear that with a generative neural network as a regularizer, the proposed method effectively mitigates the noise amplification incurred by the ill-conditioning issue associated with the low-rank and subspace reconstruction. 

\subsubsection{Impact of SNR} 
\begin{figure*}[!t]
\centerline{\includegraphics[width=\textwidth]{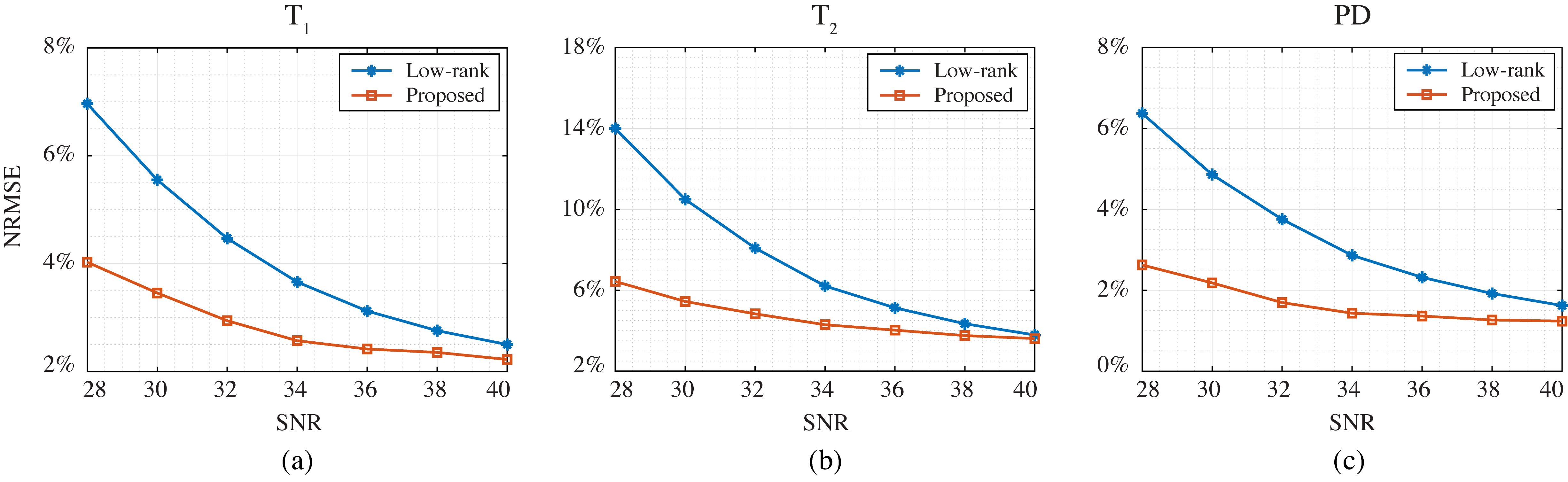}}
\caption{The NRMSE for the reconstruction of (a) $T_1$, (b) $T_2$, and (c) proton density maps from the low-rank and subspace reconstruction and the proposed method at different SNRs.}
\label{fig:nrmse_snr}
\end{figure*}

We also evaluated the performance of the proposed method at different SNR levels ranging from $28~$dB to $40~$dB at the acquisition length $M = 500$. Fig.~\ref{fig:nrmse_snr} shows the NRMSE of the reconstructed $T_1$, $T_2$, and proton density maps. As can be seen, the proposed method outperforms the low-rank and subspace reconstruction at different noise levels, and the improvement is more significant as the SNR becomes low. Again, this illustrates the effectiveness of incorporating a deep generative neural network as a regularizer for the low-rank and subspace reconstruction.  

\subsection{In Vivo Results}
We evaluated the performance of the proposed method for in vivo MR Fingerprinting experiments. Specifically, we conducted in vivo experiments on a 3T Siemens Magnetom Prisma scanner (Siemens Medical Solutions, Erlangen, Germany) equipped with a 20-channel receiver coil. A healthy volunteer was scanned using the IR-FISP sequence \cite{2015_MRM_Jiang} with the approval from the Institutional Review Board. For in vivo  experiments, we used the same set of acquisition parameters and spiral trajectories as in \cite{2015_MRM_Jiang}, and other relevant imaging parameters include FOV$ \ =300 \times 300$ mm$^2$, matrix size$ \ =256 \times 256$, and slice thickness $ \ = 5$ mm. 

Apart from the undersampled MR Fingerprinting experiments, we also acquired a set of fully-sampled data with the acquisition length $M=1000$ following the same procedure as in \cite{2016_TMI_Zhao}, and then estimated $T_{1}$, $T_{2}$, and proton density maps as our references. Moreover, we conducted a calibration scan using a vendor-provided gradient echo sequence to estimate the coil sensitivities maps. 

We performed the low-rank and subspace reconstruction and the proposed method  for in vivo data. For the above two methods, we used the same rank value (i.e., $L = 8$). In addition, we used the same dictionary for subspace estimation and dictionary matching as described in the simulations for both methods. For the proposed method, we used the same neural network architecture as described in the simulations, and manually selected the penalty parameters for optimized performance. 

\begin{figure}[!t]
\centerline{\includegraphics[width=\columnwidth]{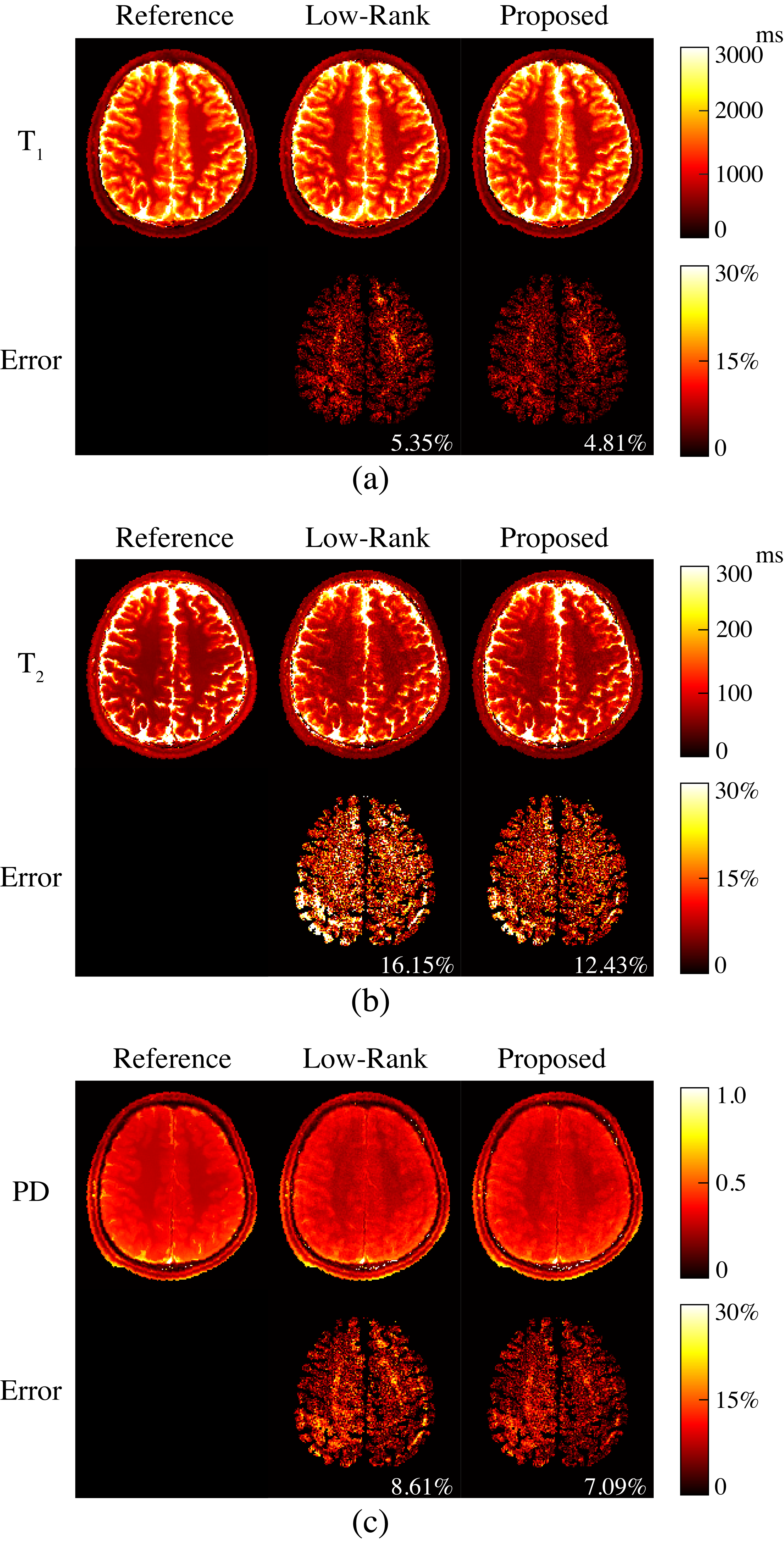}}
\caption{In vivo results from the low-rank and subspace reconstruction and the proposed method with the acquisition length $M=500$. (a) $T_{1}$ maps. (b) $T_{2}$ maps. (c) proton density maps. Note that the overall NRMSE is labeled at the right corner of each error map.}
\label{fig:params_maps_invivo}
\end{figure}

Fig.~\ref{fig:params_maps_invivo} shows the results from the low-rank and subspace reconstruction and the proposed method at the acquisition length $M = 500$. As can be seen, the proposed method outperforms the low-rank and subspace reconstruction for the $T_1$, $T_2$, and proton density maps, which is consistent with the simulation results. This further confirms the role of a deep generative neural network as a regularizer for improving the low-rank and subspace reconstruction. 

\section{Discussion}
\label{sec:discussion}
We have demonstrated the efficacy of the proposed method for utilizing an untrained deep generative prior to improve the low-rank and subspace reconstruction. Here it is worth making some further remarks on some related aspects. First, while we have illustrated the effectiveness of the proposed method utilzing a specific generative convolutional neural network, there are other alternative neural network architectures (e.g., U-net \cite{2015_MICCA_Ronneberger, 2018_CVPR_Ulyanov} or deep decoder \cite{2018_ICLR_Heckel}) that could also be adopted for the proposed method. Here it is worth pointing out that exploring the best neural network architecture for the proposed method is an interesting open problem that requires an in-depth future study. 

It is known that the use of untrained deep generative neural networks for solving inverse problems often has an overfitting issue \cite{2018_CVPR_Ulyanov}. For the proposed method, we used an early-stopping strategy to effectively mitigate this problem. Alternatively, we can incorporate additional regularizers (e.g.,  total variation \cite{2019_ICASSP_Liu} or plug-and-play \cite{2021_ICASSP_Sun} regularization) into the proposed formulation to address the issue. Although it is straightforward to extend the proposed algorithm to accommodate these regularizers, the computation time of the algorithm will inevitably increase.

We have used an untrained neural network prior to improve the low-rank and subspace reconstruction. It has been demonstrated in the early work \cite{2019_NeurIPS_Leong} that pre-training a neural network with a few examples can improve the performance of solving inverse problems over that with an untrained neural network. Here we could adopt a similar strategy for the proposed method. For example, with the use of a few training examples, we could jointly optimize the latent vector $\mathbf{z}$ and the weights of the neural network $\boldsymbol{\theta}$ in the proposed formulation, which may potentially improve the performance of the proposed method. A systematic investigation of this extension will be conducted in future work.

In addition to the architecture of the neural network, the proposed method also has a number of other hyperparameters to select, including the rank $L$ and the penalty parameters $\mu_1$ and $\mu_2$ for the algorithm. While there exist theoretical methods that could aid the selection of these hyperparameters \cite{2015_TAC_Ghadimi, 2019_ICASSP_zhang}, we can also optimize them in an empirical way, e.g., with fully-sampled reference data sets from phantom and/or in vivo experiments. The effectiveness of the latter approach has been demonstrated in this work and early methods \cite{2012_TMI_Zhao, 2015_MRM_Zhao, 2018_MRM_Zhao}. 

In this work, we have demonstrated the effectiveness of the proposed method with a conventional 2D MR Fingerprinting acquisition scheme. As a general image reconstruction method, it can also be integrated with other optimized MR Fingerprinting acquisition schemes (e.g., \cite{2019_TMI_Zhao, 2019_MRM_Asslander, 2021_PNAS_Jordan}) for further improved performance. Moreover, the proposed method can be generalized to handle volumetric MR Fingerprinting acquisitions (e.g., \cite{2016_MRM_Ye, 2017_EMBC_Zhao, 2017_NeuroImage_Liao}) to achieve faster imaging speeds. Last but not least, the proposed method can be extended to other MR parameter mapping or quantitative MRI applications. The preliminary results in our recent work \cite{2023_SSP_Lu} have demonstrated the utilities of such an extension. 

\section{Conclusion}
\label{sec:conclusion}
In this paper, we have presented a new learning-based image reconstruction method for MR Fingerprinting, which integrates low-rank and subspace modeling with a deep generative prior through an untrained neural network. The proposed method utilizes the architecture of a convolutional generative neural network to represent the spatial subspace of the low-rank model, which serves as an effective regularizer for the low-rank and subspace reconstruction. To solve the resulting optimization problem, we have described an algorithm based on variable splitting and the alternating direction method of multipliers. We have shown results from numerical simulations and in vivo experiments to demonstrate the improvement of the proposed method over the state-of-the-art low-rank and subspace reconstruction. 

\appendices


\bibliographystyle{IEEEtran}
\bibliography{refs}

\end{document}